# Incremental Analysis of Legacy Applications Using Knowledge Graphs for Application Modernization

Saravanan Krishnan
Alex Mathai
sarkris5@in.ibm.com
alex.mathai1@ibm.com
IBM Research - India

Amith Singhee
Atul Kumar
Shivali Agarwal
asinghee@in.ibm.com
kumar.atul@in.ibm.com
shivaaga@in.ibm.com
IBM Research - India

Keerthi Narayan Raghunath
David Wenk
keerag09@in.ibm.com
dwenk@us.ibm.com
IBM Consulting

## ABSTRACT

Industries such as banking, telecom, and airlines - often have large software systems that are several decades old. Many of these systems are written in old programming languages such as COBOL, PL/1, Assembler, etc. In many cases, the documentation is not updated, and those who developed/designed these systems are no longer around. Understanding these systems for either modernization or even regular maintenance has been a challenge. An extensive application may have *natural* boundaries based on its code dependencies and architecture. There are also other *logical* boundaries in an enterprise setting driven by business functions, data domains, etc. Due to these complications, the system architects generally plan their modernization across these logical boundaries in *parts*, thereby adopting an *incremental* approach for the modernization journey of the entire system.

In this work, we present a software system analysis tool that allows a subject matter expert (SME) or system architect to analyze a large software system incrementally. We analyze the source code and other artifacts (such as data schema) to create a knowledge graph using a customizable ontology/schema. Entities and relations in our ontology can be defined for any combination of programming languages and platforms. Using this knowledge graph, the analyst can then define logical boundaries around dependent Entities (e.g. Programs, Transactions, Database Tables etc.). Our tool then presents different views showcasing the dependencies from the newly defined boundary to/from the other logical groups of the system. This exercise is repeated interactively to 1) Identify the Entities and groupings of interest for a modernization task and 2) Understand how a change in one part of the system may affect the other parts. To validate the efficacy of our tool, we provide an initial study of our system on two client applications.

## 1 INTRODUCTION

While analyzing legacy applications, some common modernization patterns include: 1. *Portfolio assessment* - to create a roadmap for modernization. 2. *Decomposing monoliths* - to systemically pull out microservices using the strangler pattern [2]. 3. *Business function isolation* - to scope the functionality of an entire business process or a sub-process.

To achieve the above, architects often use strategies like retire, retain, re-factor/re-architect, re-host, re-platform and re-purchase [5]. We note that *irrespective* of the strategy used, the legacy application has to go through a critical process spanning three long phases - *discovery*, *design*, and *implementation*. To fast-track this process and drive modernization, we create a tool that 1) aids in the *discovery/understanding* phase and 2) provides a direction for the subsequent two phases in the journey. To better assess the impact of our tool, we must first understand why our approach is key to achieving the modernization scenarios. In the above scenarios, we realize a key *generalizable* problem - identifying and isolating a relevant portion of a codebase in a step-by-step manner. We address this problem with our approach 'incremental analysis'. In the incremental analysis, an SME/architect first forms a new logical group 'increment' with the most important artifacts of a specific requirement (scenario). The SME interactively identifies and gathers other related artifacts into the increment by analyzing the interactions/interfaces going across the boundaries with other logical groups. This analysis aims to come up with a collection of artifacts that we can modernize with minimal external dependencies.

To realise the benefits of incremental analysis, we create a system that enables an SME/architect to incrementally analyse an application without getting lost in the actual code/data. The key contributions of our system are: i) Representing large complex legacy applications into a knowledge graph (KG) using a

Language agnostic ontology. ii) Allowing SME input/feedback through extensions to the ontology like business function and data domain mapping. iii) Traversing the KG to compute increments and performing incremental analysis to highlight the dependencies that need to be resolved between the modernized increment and the remaining legacy artifacts. iv) Providing the KG as a base input to analysts for data science tasks like graph clustering or anomaly detection. As the KG handles the language-specific requirements, the analysts need not learn program analysis skills and can instead use that time to focus on their tasks.

## 2 SYSTEM OVERVIEW

Our deployed system has three major components that interact with each other in an end-to-end pipeline. These components are 1) *Code Discovery and Knowledge Graph Construction*, 2) *Increment Creation via Neighbourhood Detection* and 3) *Incremental Analysis*. We now proceed to explain the end-to-end flow of our tool. We begin with the input to our system i.e. legacy source code that needs to be modernized. We then run our static analyzer tool on this source code and save our analysis into an MS-SQL database. We can construct a graph representation for the input source code by combining this static analysis and a custom-made knowledge graph ontology. After saving this representation into a Neo4j [4] graph database, we run inference algorithms to create increments and subsequently complete incremental analysis to generate the corresponding insights. This entire process is detailed in Figure 1.

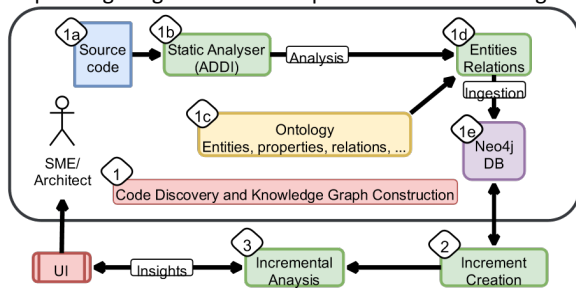

Figure 1: System Overview.

We briefly describe the three major components of our system in the following sections.

### 2.1 Code Discovery and Knowledge Graph Construction

The first step in our pipeline is to perform code discovery using static analysis. We leverage ADDI [3] as our static analyzer tool. ADDI supports numerous mainframe languages including Cobol, PL1, Assembler, and others. Using ADDI's static analysis as a foundation, we partnered with multiple SMEs to create a knowledge graph rich in semantics pertaining to legacy applications. With their domain knowledge, we developed a well-defined entity-relationship model that opens up the scope for new insights.

While defining this ontology, we pay special attention to making sure that it is language agnostic and extremely extensible. We depict a simplified version of our ontology in Figure 2. Along with entity nodes and relationship edges, we also store *attributes* for each node and edge. For example, we store node attributes like *programming language* and *Lines-of-code* for program Entities and edge attributes like *CRUD* operations on relationships. We also provide the capability for an SME to *logically* group artifacts as per domain knowledge of the codebase. We depict this in Figure 2 by the blue dotted lines (i.e. *HAS* relationship) connecting the *application* node to all it's Entities. Unlike the other relationships directly derived from the legacy code, the logical edges are subjective and can be changed as the SME is able to amass more knowledge. Using the ADDI output and the defined ontology, our ingestion component generates the knowledge graph into the Neo4j database. We will then use this database to infer insights during our next stage - *increment creation via neighbourhood detection*.

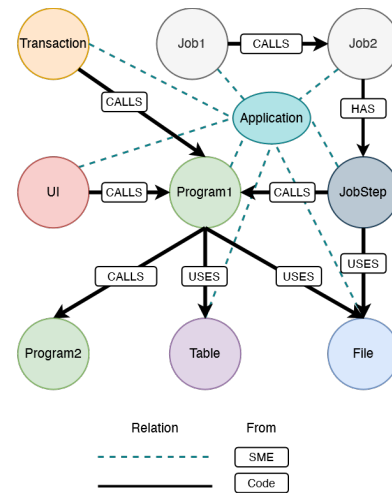

Figure 2: Sample Ontology

## 2.2 Increment Creation via Neighbourhood Detection

The discovery phase of the application modernization journey is a very time-consuming and slow process. This phase often involves a team of SMEs struggling to understand a large codebase (developed over decades) within a short stipulated time frame (few months to a year). In these situations, an SME needs to focus only on portions of the codebase relevant to the final agenda. In other words, for an SME to utilize time effectively, he/she must be able to create a boundary around *important artifacts* and successfully *ignore* the noise that lies outside this boundary.

To aid in this process, we introduce the concept of an *increment*. We define an *increment* to be a bounded scope containing *seed*

## 2.3 Incremental Analysis

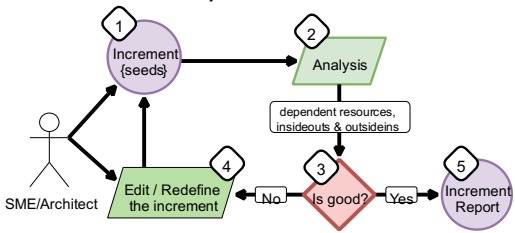

artifacts and their *neighbourhood* dependencies. As shown in Figure 3, an SME starts with *seed artifact(s)*. The seed artifacts are often something that piques the SME's interest. A seed could be of any artifact type - programs, transactions, tables, files, or others. Once the seed artifacts have been identified, we traverse the knowledge graph to gather related artifacts closely interacting with the seeds. These close artifacts denote the *neighbourhood* dependencies of the seeds. We then create a boundary around the seeds and their neighbourhoods and refer to this collection as the *increment*. We also track important metrics for each increment. Each artifact in an increment contributes to the value of the aggregate increment metric. For example, programs of an increment contribute their lines of code (LOC) and Cyclomatic complexity (Cyclo) values to the increment's aggregate LOC/Cyclometric. These metrics help the user to bring in some preferences when creating increments.

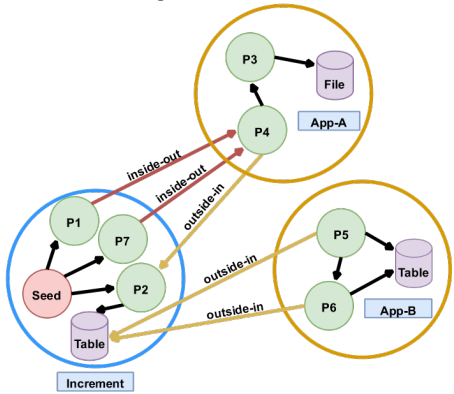

Figure 3: Incremental Analysis Flow.

Incremental analysis is a process of systematically separating a part of a codebase for a specific requirement. To understand an increment's impacts or dependencies on the rest of the codebase, we specifically look for incoming and outgoing edges from the increment's boundary. In Figure 4, we detail a simple example of incremental analysis. After choosing the seed, finding the neighbourhood, and creating the increment boundary - we highlight the outgoing red edges and the incoming yellow edges. These edges are referred to as *inside-out* and *outside-in* respectively. We can pinpoint the integration points for the new modernized code and the rest of the codebase by focusing solely on these edges.

Figure 4: Incremental Analysis.

After analysing these edges, the SME may be interested in adding or removing artifacts from the increment. As shown in Figure 4, the SME can pull program *P4* from *App-A* and add it to the increment. This move will reduce the number of dependency links. The SME will then re-analyze all the inside-out and outside-in edges and continue many iterations until the result is satisfactory. This process is also depicted in Figure 3, from steps (*Is good ?*) through (*Edit/redefine the increment*). Once the increment is finalised, the SME can begin modernizing the legacy code within the increment.

All integration points between the modernized code and the remaining codebase are made apparent to the SME with the help of the inside-out and outside-in edges. With this, it is possible to leverage the modernized code and simultaneously maintain a coexistence with the remaining code artifacts.

It is important to differentiate the benefits of incremental analysis over and above the insights derived from call graphs. The call graph Information that we get using static analysis will include every possible relationship. Considering every relationship will slow down our understanding of - how a group of artifacts as a

 whole interacts with the rest of the codebase. Our analysis uses various logical boundaries (applications and increments) that help the user focus on the critical relations/dependencies that *cross* the logical boundary.

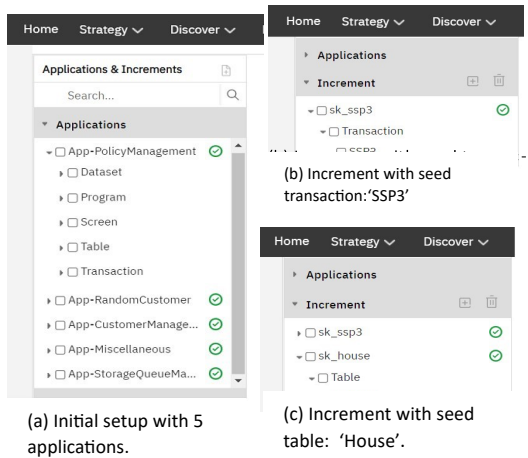

(a) Initial setup with 5 applications.

(b) Increment with seed transaction:'SSP3'

(c) Increment with seed table: 'House'.

Figure 5: Application Mapping and Increment Creation

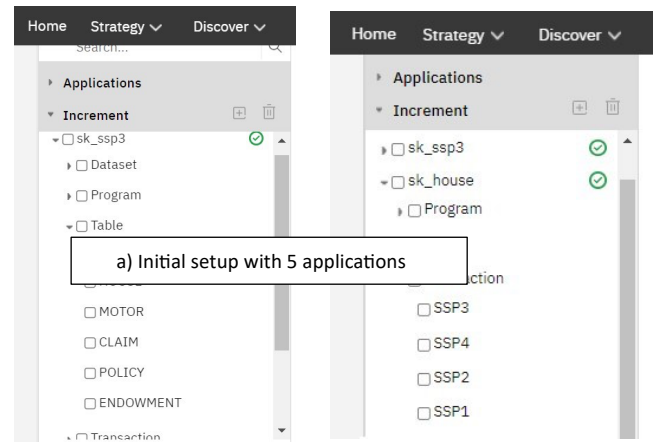

(a) Increment with seed transaction: 'SSP3'

(b) Increment with seed table: 'House' expanded

Figure 6: Increment Expansion

## 3 SYSTEM DEMONSTRATION

We demonstrate our system using a publicly available standard mainframe application - *General insurance applica2on (GENAPP)* [1]. Even though we demonstrate our system on a mainframe application, it is *not* limited to mainframes. As the approach and the ontology are extensible, our system is agnostic. To focus the demonstration on the actual incremental analysis, let us assume the following - GENAPP source code is analysed by the static analyser (ADDI), Entities and their relationships are mapped with the ontology, and the data is available in the Neo4j database. At the ingestion stage, all the artifacts of GENAPP are grouped under five applications - Customer Management, Policy Management, Random Customer, Storage Queue Management, and Miscellaneous. They appear in the tool (Figure 5a) under the section 'Applications'.

Let us consider the case where we would like to Identify the sub-application related to 'House insurance policy' in GENAPP that we need to take through in one of the modernization strategies. Before doing the analysis, we need to have a very high-level understanding of the transactions or data resources (*not* the code implementation) present in the application (e.g. the transactions 'SSP1', 'SSP2', 'SSP3' and 'SSP4' are related to the 'Motor', 'Endowment', 'House' and 'Commercial' policies respectively. There are also separate tables for these policies with the same name.)

Demo scenario 1: If we just know the specific transaction 'SSP3' associated to 'House policy', then create an increment with this transaction (Figure 5b) and trigger the analysis. This results in first expanding the increment by pulling in all neighbourhood dependencies of 'SSP3' and showing the inside-outs and outside-ins. The resultant increment has one seed transaction, 13 programs and 6 tables (House, Commercial, Motor, Endowment, Claim, Policy) (shown in Figure 6a). While we are expecting only one table 'House', we get five more. This is due to the fact that the programs interacting with 'House' also interact with the other tables.

Demo scenario 2: If we just know the specific table 'House' associated to 'House policy', then create an increment with this table and trigger the analysis. This first results in expanding the increment by pulling in all related artifacts of the table 'house' and showing the inside-outs and outside-ins. The resultant increment has one seeded table, 13 programs and the 4 transactions (SSP1, SSP2, SSP3, SSP4). While we are expecting only one transaction 'SSP3', we get three more that are associated to the other policies. This could be due to the fact that the programs are common to interact all the tables and all the transaction are using these common programs.

3.5$k$ artifacts. A video of our system demonstration is available at: github.com/IBM/AppModDemos/CAdemos.

| | Programs | Jobs | Transactions | Files | Tables | ... | Total |
|---|---|---|---|---|---|---|---|
| ClientA | 3.5k | 12k | 680 | 95K | 1.1K | | 513k |
| ClientB | 5k | 1.5k | 2.3k | 48k | 339 | | 1000k |

Figure 8: Details of two client applications

## 4 CONCLUSIONS AND FUTURE WORK

In this work, we introduced the concept of incremental analysis and explained its benefits when modernizing applications. We showcased our knowledge graph and emphasized that it is language agnostic and allows additions beyond static analysis - like business functions and data domains. Finally, we demonstrated our system on GENAPP and highlighted our insights. While our current tool leverages static analysis, our future works include - understanding the data of the application (ex. tables) and insights extraction from operational logs.

In both the above scenarios, we see that the programs are common to interact with all the tables and hence if we need to create separate micro-service for 'House policy', we need to apply 're-factor' strategy to decompose those common programs. Let us consider the case where we need to retire the functionality 'random customer'. Assume that we know that the related transaction is 'LGCF' and hence create an increment with it. The analysis results in expanding the increment with only one program 'LGICVS01'. From Figure 7, we can observe the following points : (1) 'LGICVS01' is interacting the file 'GENAPP.GENAPP.KSDSCUST' with just read operation; (2) same program is interacting to the message queue 'GENACNTL' to just write the log message; (3) there aren't any other program that calls program 'LGICVS01'. With the above three points, we can be sure of the transaction 'LGCF' and the associated program 'LGICVS01' are not dependent for the rest of GENAPP and hence it can be retired safely without any changes.

Figure 7: Increment with seed transaction: 'LGCF'

System performance : An initial study of applying the above demo scenarios on GenApp results in an increase in productivity of 20%. In Figure 8, we show the number of key artifacts present in the two client applications that we are currently analysing. Our system takes roughly 20 seconds to process an increment having

REFERENCES
[1] IBM CICS. 2020. *General insurance application (GENAPP) for IBM CICS TS*. Retrieved August 12, 2021 from https://www.ibm.com/support/pages/node/310731
[2] IBM. 2020. *Break the monolith: Chunking strategy and the Strangler pattern*. Retrieved August 12, 2021 from https://www.ibm.com/garage/method/prac9ces/code/chunking-strategy-strangler-pattern/
[3] IBM. 2021. *IBM Application Discovery and Delivery Intelligence - Overview*. Retrieved August 12, 2021 from https://www.ibm.com/in-en/products/appdiscoveryand-delivery-intelligence
[4] Neo4j. 2020. *Neo4j Main page*. Retrieved August 12, 2021 from https://neo4j.com/
[5] Amazon Web Services. 2020. *Application Migration Strategies*. Retrieved August 12, 2021 from https://docs.aws.amazon.com/whitepapers/latest/aws-migration-whitepaper/welcome.html